\documentstyle[twocolumn,aps,prb,epsfig]{revtex}
\begin{document}
\draft
\title{Guiding center picture of magnetoresistance oscillations in rectangular
  superlattices}
\date{\today}
\author{Rolf R. Gerhardts and Stephan D. M. Zwerschke}
\address{
Max-Planck-Institut f\"ur Festk\"orperforschung, Heisenbergstrasse 1,
D-70569 Stuttgart, Germany}

\twocolumn[
\setlength{\columnwidth}{18cm}%
\csname@twocolumnfalse\endcsname 
\maketitle
\begin{abstract}
We calculate the magneto-resistivities of a two-dimensional electron gas
subjected to a lateral superlattice (LSL) of rectangular symmetry within the
guiding-center picture, which approximates the classical electron motion as a
rapid cyclotron motion around a slowly drifting guiding center. We explicitly
evaluate the velocity auto-correlation function along the trajectories of the
guiding centers, which are equipotentials of a magnetic-field dependent
effective LSL potential.
The existence of closed equipotentials may lead to a suppression of the 
commensurability oscillations, if the mean free path and the LSL 
modulation potential are large enough.  We present numerical and
analytical results for this suppression, which allow, in contrast to previous
quantum arguments, a classical explanation of similar suppression effects
observed experimentally on square-symmetric LSL. Furthermore, for rectangular
LSLs of lower symmetry  they lead us  to predict  a strongly anisotropic
resistance tensor, with high- and low-resistance directions  which can be
interchanged by tuning the externally applied magnetic field. 

\end{abstract}
]

\section{Introduction}
Pronounced commensurability oscillations of the magnetoresistance of a
two-dimensional 
electron gas (2D EG) subjected to a perpendicular magnetic field and a
lateral superlattice,  now also known as Weiss oscillations (WO),
have first been observed on systems with a periodic modulation in one
 direction (1D). \cite{Weiss89:179,Gerhardts89:1173,Winkler89:1177}
Subsequent work on systems with  a 2D  lateral superlattice (LSL) showed that
the modulation in the second direction tends to suppress the commensurability
oscillations.
\cite{Weiss90:88,Fang90:10171,Gerhardts91:5192,Lorke91:3447,Weiss92:314}
The  WO observed on samples with a 1D LSL, and their suppression in samples
with a 2D LSL, were first explained by quantum mechanically.
A 1D modulation broadens the Landau levels into bands of oscillatory width,
with finite  group velocity. This leads, in
addition to the scattering induced magnetoconductivity (the so called
``scattering conductivity''), to a ``band conductivity'' which vanishes if the
Landau bands become flat. \cite{Gerhardts89:1173,Winkler89:1177,Zhang90:12850}
A modulation in the second lateral direction splits these Landau bands into
narrow subbands with  small group velocities,
and, as a function of the magnetic flux per unit cell of the 2D
LSL,  a self-similar energy spectrum  (``Hofstadter's butterfly'') results.
\cite{Hofstadter76:2239,Pfannkuche92:12606}
 It has been argued that
this subband splitting leads to a suppression of the band conductivity,
if the modulation-induced width of the Landau bands is sufficiently
large, and  collision broadening effects are sufficiently weak. 
 \cite{Gerhardts91:5192,Pfannkuche92:12606}
Experiments on samples with a weak 2D modulation and not too high mobility
show indeed commensurability oscillations very similar to those observed in 1D
LSLs, which are suppressed (and changed in
character) with increasing modulation strength and mobility.
\cite{Lorke91:3447,Weiss92:314}

Soon after their discovery,  Beenakker \cite{Beenakker89:2020} explained the
most prominent of the WO  in an electrostatically defined 1D
LSL   classically, as resulting from an
oscillatory ${\bf E \times B}$ drift of the guiding centers (GCs) of cyclotron
orbits, where the GC
velocity  plays the role of the group velocity in the
 quantum treatment. \cite{Zwerschke99:5536} The GC picture can be 
justified for weak modulations and intermediate strengths of the
applied magnetic field. \cite{Gerhardts92:3449,Menne98:1707}  It 
 has been used to calculate  the resistivity
for different analytical forms of electrostatic, magnetostatic and mixed 
modulations defining 1D or 2D LSLs. \cite{Gerhardts96:11064} A clear
classical picture for the suppression of the WO in 2D LSLs
has, however, not been developed for nearly a decade, although direct
numerical  evaluations of  the diffusion tensor on the basis
of classical ballistic models indicated such a suppression.
 \cite{Lorke91:3447,Schmidt93:13007}

 Recently Grant {\em et al.} \cite{Grant00:13127} emphasized
that the GCs  move along the equipotential lines of an effective
potential, determined by an average of the 
modulation fields over unperturbed cyclotron orbits.
\cite{Schmidt93:13007} They 
argued that, in a 2D LSL of square symmetry, these equipotentials are closed,
and that therefore the GC velocity averages to zero, resulting in
a suppression of the WO, and they confirmed this conjecture by ballistic model
calculation.  Recent experiments
demonstrate also, that an asymmetric 2D modulation leads to
much stronger commensurability oscillations than a square-symmetric one does.
\cite{Chowdhury00:R4821}
While Grant {\em et al.} \cite{Grant00:13127} employed the GC
picture to make the suppression of WO in 2D LSL plausible,
they did not really use it as basis of their calculation. Moreover, they
considered a strong modulation, so that their results are not directly
comparable with previous predictions for weak modulation.
\cite{Gerhardts96:11064} A consistent evaluation of the GC
approach, which is known to yield  a very 
simple and intuitive picture of the WO in 1D LSLs, is so far
not available for the case of 2D LSLs. 
The aim of the present work is to fill this gap.

  In Sect. \ref{gc-pict} we discuss the  GC approach and its
 limitations.  Based on a numerical evaluation
 of the diffusion tensor, we present in  Sect.\ \ref{result-exam}
simple analytical results for the
conductivities of square and rectangular LSLs defined by harmonic electric and
magnetic modulation fields. These results depend only on one (square LSL)
respectively two (rectangular LSL) parameters, which are determined by the
(seven) modulation model parameters, the mean free path, and the average
magnetic field. Illustrative examples are given, including a strongly
anisotropic case in which the applied magnetic field can interchange the
directions of high and low resistance. Mathematical details are given in two
appendices.

\section{The guiding center picture} \label{gc-pict}

\subsection{Heuristic definition}
We consider a 2D EG in the $x$-$y$ plane subjected to 
a strong homogeneous, perpendicular magnetic field ${\bf B}_0=(0,0,B_0)$ and a
LSL defined by weak electric and magnetic modulation fields. 
The classical magnetoconductivity of this system can be calculated
\cite{Gerhardts92:3449} 
from the motion of electrons at the Fermi energy $E_F=(m/2)v_F ^2 $. Within the
GC picture, this is assumed to be the superposition
\begin{equation} \label{guid-cyc}
{\bf r}(t)={\bf r}_{\mathrm gc}(t)+{\bf r}_{\mathrm cyc}(t)
\end{equation}
 of a rapid  cyclotron motion 
${\bf r}_{\mathrm cyc}(t)=R (\sin \alpha,-\cos \alpha)$ around a slowly moving
guiding 
center ${\bf r}_{\mathrm gc}(t)$, where $\alpha(t)=\omega_0 t+ \alpha_0$
describes a 
uniform circular motion with  cyclotron frequency $\omega_0=e B_0/(mc)$ and
 radius $R=v_F/\omega_0$. This is obviously correct in the absence of
 modulation fields, where the position of the GC ${\bf r}_{\mathrm
   gc}(t)$ is a constant of motion, and in 
the presence of a homogeneous in-plane electric field ${\bf E \perp B}_0$,
where the GC moves with the constant drift velocity 
$\dot{{\bf r}}_{\mathrm gc}= c ({\bf E \times B}_0)/B_0^2$. 
For a perturbation by a position-dependent in-plane  electric field 
${\bf E  =\nabla}V({\bf r})/e$ or perpendicular magnetic field
${\bf B}_m=(0,0,B_m({\bf r}))$, the GC picture is only
approximately valid, and several definitions of a ``guiding center'' are
possible, which become equivalent in the limit of small perturbations.

 A reasonable candidate  is the center of the circle of curvature
at the point ${\bf r}(t)$. Taking the energy conservation $(m/2) v^2+V({\bf
  r})=E_F$ into account and writing the velocity as ${\bf \dot r=v}=v({\bf r})
(\cos \varphi, \sin \varphi,0)$ with $v({\bf r})=v_F [1-V({\bf
  r})/E_F]^{1/2}$, this center is given by \cite{Gerhardts96:11064} 
\begin{equation} \label{cccurve}
{\bf r}_M={\bf r}+{\bf e_z \times v}/(\omega_0+ \omega_{\mathrm mod}) \, ,
\end{equation}
where $\omega_{\mathrm mod}=\omega_m +[{\bf e_z}
\times ({\bf v}/v)] \cdot {\bf \nabla} v({\bf r})$ with
$ \omega_m ({\bf r}) =e B_m({\bf r})/(mc)$.

To lowest order in the modulation strength one may neglect the modulation
effect $\omega_{\mathrm mod}$
in the denominator of Eq.~(\ref{cccurve}). Then Newton's equation 
$m {\bf \dot v} =-e [{\bf E}+ ({\bf v}/c)\times ({\bf B}_0+{\bf B}_m)]$ yields
the time derivatives \cite{Gerhardts96:11064} 
\begin{equation} \label{driftfuzz}
\dot x_M=c\, \frac{E_y}{B_0} -v_x\, \frac{\omega_m}{\omega_0} \, , \quad
\dot y_M=-c\, \frac{E_x}{B_0} -v_y\, \frac{\omega_m}{\omega_0}\, .
\end{equation}
In the spirit of the GC picture we may average
Eq.~(\ref{driftfuzz}) over the rapid cyclotron motion, i.e., we approximate
${\bf v} \approx {\bf \dot r}_{\mathrm cyc}$, replace ${\bf r}(t)={\bf
  r}_{\mathrm   gc}(t)+{\bf r}_{\mathrm cyc}(t)$ in the arguments of ${\bf E}$
and $\omega_m$, and take 
the average with respect to $\alpha = \omega_0 t$ over one period. We assume
that $V({\bf r})$ and $\omega_m({\bf r})$ are periodic, with vanishing average
values, on the same rectangular
lattice with lattice constants $a_x=2\pi/K_x$ and $a_y=2 \pi /K_y$,
\begin{equation} \label{fourier}
V({\bf r})=\sum_{\bf q \neq 0} V_{\bf q}\, e^{i \bf q \cdot r} \, , 
\quad
\omega_m({\bf r})=\sum_{\bf q \neq 0}\omega_{\bf q}\, e^{i \bf q \cdot r} \, ,
\end{equation}
where  ${\mathbf{q}}=(n_x K_x, n_y K_y)$. The  averages over the cyclotron
motion can be performed for each Fourier component separately.
\cite{Gerhardts96:11064} The result of this approximation is the equation of
motion for the GC,
\begin{equation}  \label{guid-drift}
{\bf \dot r}_{\mathrm{gc}} =
{\mathbf{v}}_{\mathrm{gc}} = - \frac{1}{m \omega_0} \, {\mathbf{e_z} \times
    \mathbf{\nabla}} V_{\mathrm{eff}} (\mathbf{r}_{\mathrm{gc}})   \, ,
\end{equation}
where the effective potential
$V_{\mathrm{eff}} ({\mathbf{r}}) =\sum_{\mathbf{q} \neq \mathbf{0}}
    e^{i\mathbf{q \cdot r}} \, V^{\mathrm{eff}}_{\mathbf{q}}$,
is determined by \cite{Gerhardts96:11064,Grant00:13127} 
\begin{equation} \label{veff-coef}
 V^{\mathrm{eff}}_{\mathbf{q}} = V_{\mathbf{q}} J_0(qR)+ \frac{mv_F}{q}\,
 \omega_{\mathbf{q}}  J_1(qR) \, ,
\end{equation}
with Bessel functions $J_0$ and $J_1$. 
According to
Eq.~(\ref{guid-drift}),  the GC moves along the equipotential
lines of 
the effective potential $ V_{\mathrm{eff}} (\mathbf{r}_{\mathrm{gc}})$. Note
that, in Eq.~(\ref{guid-drift}), we have identified the GC with
the average value (over one cyclotron cycle) of ${\bf r}_M(t)$ as defined in
Eq.~(\ref{cccurve}), and that we have taken into account only terms in lowest
order of $\omega_{\rm mod}/\omega_0 \ll 1$. 

\subsection{Examples and limitations}
We have, for a large number of examples, integrated
Newton's equation numerically to obtain the exact trajectories ${\bf r}(t)$
and ${\bf r}_M(t)$ as defined in
Eq.~(\ref{cccurve}), and also with the approximation $\omega_{\mathrm  mod} /
\omega_0 \rightarrow 0$. We found that in all cases with sufficiently small 
modulations (roughly $|\omega_{\mathrm  mod}| \lesssim 0.2\,
\omega_0$)  the average of ${\bf r}_M(t)$ and of its approximation 
for $\omega_{\rm mod} \rightarrow 0$ are
practically identical. Moreover, these averages  agree with the 
average of the exact trajectory ${\bf r}(t)$ over the cyclotron motion, which
we have calculated as 
\begin{equation} \label{avertraj}
{\bf \bar r}(t)=\frac{1}{T_+(t)-T_-(t)} \int_{T_-(t)}^{T_+(t)} dt'\,
{\bf  r}(t') \, , 
\end{equation}
where $ \varphi(T_{\pm})= \varphi(t) \pm \pi$, and $ \varphi(t)$  defines the
direction of the velocity at time $t$,  ${\bf \dot r}(t)=v({\bf r}(t))
(\cos \varphi(t), \sin   \varphi(t),0 )$. 
This rather complicated definition
of the time average seems necessary, since in the 2D LSL the velocity vector 
${\bf \dot r}(t)$ is not an exactly  periodic function of time, in contrast to
the case of drifting orbits in a 1D LSL. \cite{Zwerschke99:5536}

\begin{figure}[h]  \centering
\noindent  
\includegraphics[width=1.15\linewidth]{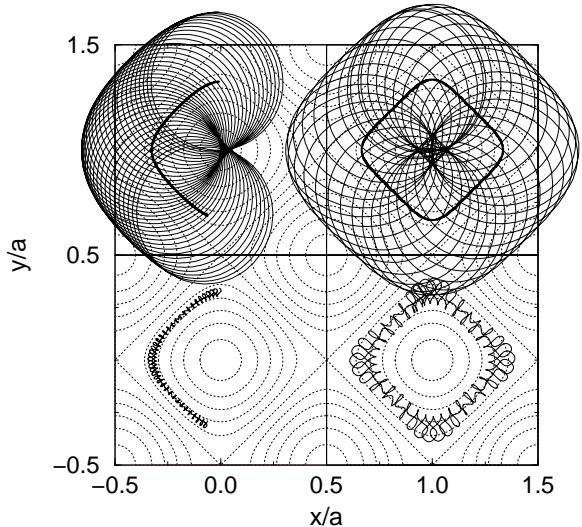}
\caption{\small Parts of rosette-like orbits (about 39 cyclotron cycles, same
  initial conditions) in the electric modulation potential 
$V(x,y)=\varepsilon \,E_F [\cos qx + \cos qy]$, with $q=2\pi /a$ and $qR=2$,
for $\varepsilon =0.05$ (left) and $\varepsilon =0.15$ (right). Thick solid
lines show the corresponding  ${\bf \bar r}(t)$ as
 defined by  Eq.~(\protect\ref{avertraj}), thin dotted lines the
equipotentials of $V(x,y)$. The corresponding ${\bf r}_M(t)$,
Eq.~(\protect\ref{cccurve}) with $\omega_{\mathrm mod} = 0$, are  shifted
downwards by one lattice period.
\label{fig:rosett}}
\end{figure}

\begin{figure}[h]  \centering
\includegraphics[width=1.2\linewidth]{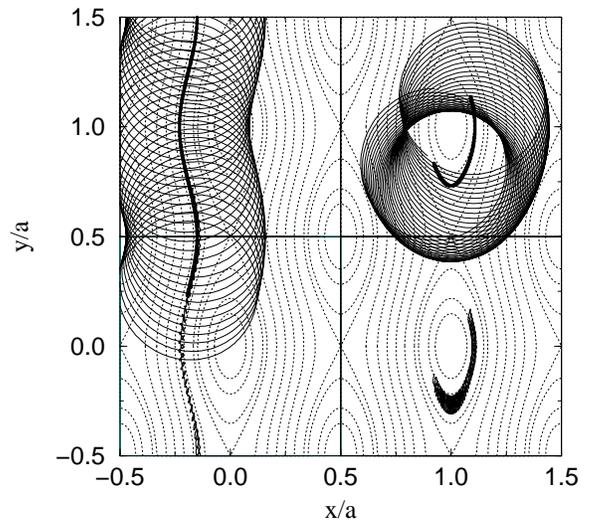}
\caption{\small Thirty-eight cyclotron cycles of a drifting (left) and of a
  localized   (right) orbit in the electric modulation potential 
$V(x,y)=0.1\,E_F [\cos qx + 0.25\, \cos qy]$ (equipotentials indicated by thin
dotted lines), with $q=2\pi /a$ and $qR=2$. Thick  solid lines:  GC
trajectories as defined by  Eq.~(\protect\ref{avertraj}).
 The rapidly fluctuating trajectories of  ${\bf r}_M(t)$,
 Eq.~(\protect\ref{cccurve}) 
with $\omega_{\mathrm mod} = 0$, are shifted downwards
 by one period.
\label{fig:bahnen}}
\end{figure}

As a typical example we show in Fig.~\ref{fig:rosett}, for two different
modulation strengths of a square-symmetric electrical modulation,
rosette-like orbits together with   ${\bf r}_M(t)$ (in the limit $\omega_{\rm
  mod}\rightarrow 0$) and ${\bf \bar r}(t)$  defined by Eqs.~(\ref{cccurve})
and (\ref{avertraj}), respectively. For weak modulation, both definitions
yield trajectories 
close to equipotentials, as expected from Eq.~(\ref{guid-drift}). However,
${\bf r}_M(t)$ exhibits  rapid fluctuations around its
equipotential, with an amplitude that increases with the modulation strength.
Using the cyclotron motion as a reference, we see from Fig.~\ref{fig:rosett}
that the velocity of the GC motion increases (essentially linearly)
with increasing modulation strength.
If the modulation has only a rectangular instead of a square symmetry, the
GCs may follow either closed (localized) or open (drifting)
equipotentials. 
A typical example with an electrostatically defined LSL is shown in
Fig.~\ref{fig:bahnen}. 

We want to mention that there are situations in which the GC
picture works, but Eq.~(\ref{guid-drift}) does not. 
This is, e.g., the case, if due to the
Bessel functions in Eq.~(\ref{veff-coef}) the effective potential vanishes for
finite modulation, and therefore the first order approximation
Eq.~(\ref{guid-drift}) fails.
 For the square-symmetric harmonic electric modulation
considered in Fig.~\ref{fig:rosett}, this happens for $J_0(qR)=0$ (``electric
flat-band condition'').
Then, besides orbits with GCs moving around potential maxima [${\bf
  r}=(ma, na)$] and minima [${\bf r}=(2m+1, 2n+1)a/2$], there are also orbits
  with 
GCs moving around saddle points at [${\bf r}=(2m, 2n+1)a/2$] and
[${\bf r}=(2m+1, 2n)a/2$], which are not described by Eq.~(\ref{guid-drift}).
The approximation Eq.~(\ref{guid-drift}) becomes also poorer with
increasing modulation strength. Thus, we  see from the right thick line in
Fig.~\ref{fig:rosett} that the GC   deviates
characteristically from  the equipotential trajectory
predicted by Eq.~(\ref{guid-drift}).

The failure of Eq.~(\ref{guid-drift}) near the ``flat-band conditions'' is
 not very important for the calculation of the conductivity, since
there the GC drift is anyway slow and thus contributes little to the
conductivity. 
There are, however, natural limitations of the GC picture. Of
course, if the 
(average) magnetic field $B_0$ becomes too strong, the classical approach
fails and Landau quantization effects must be taken into account. If $B_0$
is zero or very small, ``channeled orbits'' occur similar to the case of a 1D
LSL.  \cite{Zwerschke99:5536} In contrast to the 1D case, in 2D LSL one also
observes  chaotic orbits
\cite{Fleischmann92:1367,Schuster93:6843} if the modulation is 
sufficiently strong and $B_0$ is sufficiently small, $\omega_{\mathrm
  mod} \gtrsim \omega_0$. In the regime of chaotic orbits (i.e., for the model
of Fig.~\ref{fig:rosett} with $\varepsilon \sim 0.05$ for $qR  \gtrsim 7$)
 the GC picture is not useful, since it does not simplify the description of
 the electron motion.
In the following we will consider the decomposition of Eq.~(\ref{guid-cyc})
for the electron motion in a LSL and calculate the GC motion from 
Eq.~(\ref{guid-drift}). This is a good approximation if the modulation fields
defining the LSL are sufficiently weak and if the average magnetic field 
$B_0$ is sufficiently strong.
 We want to emphasize that the GC picture yields
reasonable results for the magnetoconductivity even in situations in which 
Eq.~(\ref{guid-cyc}) with ${\bf r}_{\mathrm gc}(t)$ calculated from
Eq.~(\ref{guid-drift}) does not yield a reasonable approximation for an
individual trajectory ${\bf r} (t)$  with the same initial conditions.

\subsection{Conductivity from GC motion} \label{cond-gcm}
To calculate the magnetoconductivity of a 2D EG in a LSL within the
relaxation time approximation, we use Einstein's relation $\sigma_{\mu \nu}=
D_{\mu \nu} e^2 m/(\pi \hbar^2)$ and
 the Chambers formula \cite{Chambers69:175,Gerhardts96:11064} for the
diffusion tensor $D_{\mu \nu}$, which contains the
velocity autocorrelation 
integral along a trajectory, averaged over all initial conditions
${\bf r} (0)={\bf r}_0 $, ${\bf \dot r} (0)= v({\bf r}_0)\, (\cos \varphi_0,
\sin \varphi_0)$. With the decomposition of Eq.~(\ref{guid-cyc}) this yields
three types of terms. One term, which contains only the cyclotron velocity and
must be 
averaged over the initial value $\alpha_0$,  yields the Drude
conductivity tensor. The mixed terms, containing both the cyclotron and the
GC velocity, vanish upon averaging over $\alpha_0$.
Finally, the term containing only the GC drift contribution is given by
\begin{equation} \label{diff-tens-gc}
D^ {\mathrm{gc}}_{\mu \nu} = \int_0 ^{\infty} \! dt \, e^{-t/\tau}\Big\langle
v_{\mu}^ {\mathrm{gc}}(t) v_{\nu}^ {\mathrm{gc}}(0)
\Big\rangle_{\mathrm{init}} \, , 
\end{equation}
where the average has to be taken over all possible initial positions  
${\mathbf{r}}_{\mathrm{gc}}(0)$ of GC trajectories in a unit
cell of the periodic potential.

With the dimensionless coordinates $\xi=K_x x$ and $\eta=K_y y$ and
the effective potential 
\begin{equation} \label{wxieta}
w(\xi,\eta)= \sum_{m , n} e^{i(m \xi +n \eta)} \,
V^{\mathrm{eff}}_{(m K_x, n K_y)} / V_{\mathrm{cha}} \, ,
\end{equation}
where $V_{\mathrm{cha}}$ is an energy  characteristic for $V_{\mathrm{eff}}
(\mathbf{r})$, e.g. its maximum,
 Eq.~(\ref{guid-drift}) reads
\begin{equation} \label{xi-eta-dot}
\frac{d\xi}{dt} =\Omega \, \frac{\partial w}{\partial \eta} \, , \quad 
\frac{d\eta}{dt} = - \Omega \, \frac{\partial w}{\partial \xi} \, ,
\end{equation}
with 
\begin{equation}  \label{Kap-omega}
\Omega = K_x K_y \, \frac{V_{\mathrm{cha}}}{m \omega_0} \, .
\end{equation}
Given the analytical form of the effective potential $w(\xi,\eta)$, we can use
Eqs.~(\ref{xi-eta-dot}) to calculate the diffusion tensor (\ref{diff-tens-gc})
as an integral over the equipotentials of $w(\xi,\eta)$ (see appendix A). The
results then 
depend only on the parameter $\Omega \tau$. In the following section we will
demonstrate this with a few explicit examples.

\section{Results and examples} \label{result-exam}
To keep the notation simple, we will in the following consider only
superlattices with a rectangular symmetry, with 1D and square-symmetric LSL as
limiting cases. 

\subsection{One-dimensional modulation} \label{one-d-mod}
If the periodic potential depends only on one coordinate, say $w(\xi)$,
Eq.~(\ref{xi-eta-dot}) yields $v_x=\dot{\xi}/K_x=0$, $\xi(t)=\xi(0)$, and 
$v_y= w'(\xi(0)) /K_y$, independent of time. Thus we can immediately
evaluate Eq.~(\ref{diff-tens-gc}) to obtain $D^{ \mathrm{gc}}_{xx} =
D^{ \mathrm{gc}}_{xy} = D^{ \mathrm{gc}}_{yx} =0$ and 
\begin{equation} \label{diff-t-gc1}
D^{ \mathrm{gc}}_{yy}\! = \! \frac{ \Omega^2 \tau}{ K_y^{2}} \!
\int_{-\pi}^\pi \! \frac{d\xi}{2\pi}  \Big[w'(\xi)\Big]^2 \!
\equiv  \tau \left[\frac{K_x V_{\mathrm{cha}}}{m \omega_0 }\right]^2 \!
\left\langle \! \Big[w'(\xi)\Big]^2 \! \right\rangle _{\xi}
 \, .
\end{equation}
For simple harmonic modulations, e.g., $V_{\bf q}= \delta _{{\bf q},(\pm K,0)}
V_K$ and  $\omega_{\bf q}= \delta _{{\bf q},(\pm K,0)} \omega_{\pm K}$ with
$\omega_{-K}= \omega_K^*$ in Eq.~(\ref{fourier}), one obtains from
Eq.~(\ref{veff-coef})
 $V_{\mathrm{eff}} ({\bf r})= \tilde{V}_0 \cos (Kx +\alpha)$, with
 $\tilde{V}_0=2\, |V^{\rm eff}_{(K,0)}|$. With $V_{\mathrm{cha}}= \tilde{V}_0$
 and  $w'(\xi)=-\sin (\xi+\alpha)$, 
 Eq.~(\ref{diff-t-gc1}) reproduces the known formula \cite{Gerhardts96:11064}
\begin{equation} \label{dsig-1d}
\Delta \sigma_{yy}^{\rm 1D}= \frac{e^2m}{\pi \hbar^2} \, \frac{\Omega^2
  \tau}{2K^2} \, ,
\end{equation}
where the WO result from the oscillatory behavior of 
\begin{equation} \label{Omega-har}
\Omega =\frac{K^2 v_F^2}{\omega_0} \left| \frac{V_K}{E_F}\,  J_0(KR) +
\frac{2 \omega_K}{K v_F} \, J_1(KR) \right| \, .
\end{equation}
Here a complex ratio $\omega_K /V_K$ allows to describe a phaseshift between
electric and magnetic modulation. 

\subsection{Weak 2D modulation, $\Omega \tau \ll 1$}
For a 2D superlattice potential  one obtains a similar simple
result, if the modulation amplitude (or the relaxation
time $\tau$) is sufficiently small, so that $\Omega \tau \ll 1$. Then
 we may approximate $v_{\mu} (t) \approx v_{\mu} (0)$ in
Eq.~(\ref{diff-tens-gc}), so that the $t$ integral becomes trivial, with
the result 
\begin{equation} \label{diff-t-gc-small}
D^{\mathrm{gc}}_{\mu \nu}=\frac{\sigma_{\mu} \sigma_{\nu}}{K_{\mu} K_{\nu}}
\, \frac{ 
\Omega^2 \tau}{(2\pi)^2} \int_{-\pi}^{\pi} \! d\xi \int_{-\pi}^{\pi}\! d\eta \,
w_{\bar{\mu}}\, w_{\bar{\nu}}
\end{equation}
with $\sigma_x=1$, $\sigma_y=-1$,  $w_x=\partial w/\partial \xi$, $w_y=\partial
w /\partial \eta$ and the notation $\bar{\mu}=y$ (or $x$) if $\mu=x$ (or $y$).
 The factor $(2\pi)^2$ is the area of the dimensionless unit
 cell. 

If the periodic potential is additive, $w(\xi,\eta)
=w^{(1)}(\xi)+w^{(2)}(\eta)$, the off-diagonal elements vanish, $D^
{\mathrm{gc}}_{xy} =D^{\mathrm{gc}}_{yx} =0$, and the diagonal elements
agree with those of the corresponding 1D modulations, Eq.~(\ref{diff-t-gc1}).

This weak-modulation limit $\Omega \tau\ll 1$, in which the
magnetoconductivity is independent of the nature of the GC
trajectories, has been discussed in  Ref.~\cite{Gerhardts96:11064}. 
 However, with increasing modulation strength (and larger
relaxation time, i.e.\ 
larger mean free path) the nature of the GC trajectories will
become important. For $\Omega \tau \gg 1$ the time integral will be
proportional to the average velocity along the trajectory. For closed
trajectories, this average will vanish, whereas for open equipotentials, which
may exist either in $x$ or in $y$ direction, the average may be finite.
Thus we expect that, in the limit $\Omega \tau \rightarrow \infty$, closed
equipotentials  will not contribute to the diffusion tensor, whereas the
contribution of open ones will be similar to the case of 1D
modulation.

\subsection{Square-symmetric harmonic modulation}
\begin{figure}[t]  \centering
\includegraphics[width=0.9\linewidth]{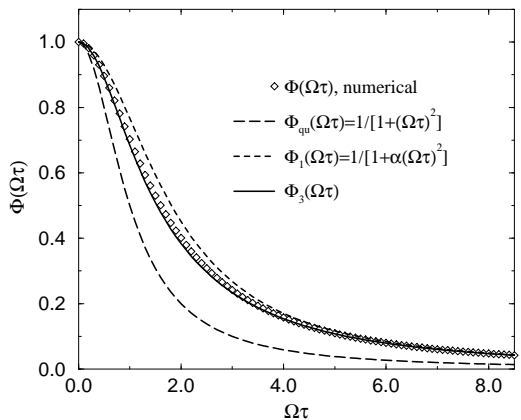}
\caption{\small Suppression of conductivity in 2D superlattice with square
  symmetry; numerical result $\Phi(\Omega \tau)$
  (open diamonds),
  quadratic approximation with cut-off energy $\epsilon_{\rm qu}=0.228$
  (long-dashed), 
   one-parameter interpolation with $\alpha=0.304$ (dashed), and  the
  three-parameter interpolation of Eq.~(\protect{\ref{tre-par-int}})
  (solid line).
\label{fig:fvonXiso}}
\end{figure}

We now consider the 2D version of the simple harmonic modulation discussed in
Sect.~\ref{one-d-mod}, i.e., assume in Eq.~(\ref{fourier})
$V_{\bf q}= (\delta _{{\bf q},(\pm K,0)}+ \delta _{{\bf q},(0,\pm K)}) V_K$
and $\omega_{\bf q}= (\delta _{{\bf q},(\pm K,0)}+ \delta _{{\bf q},(0,\pm K)})
 \omega_{\pm K}$ with $\omega_{-K}= \omega_K^*$. Then the effective potential
 has the form 
\begin{equation}  \label{cossum_squ}
 V_{\mathrm{eff}} ({\bf r})= \tilde{V}_0 [ \cos (Kx +\alpha) +
\cos (Ky +\alpha)]\, ,
\end {equation}
with an (irrelevant) phase shift $\alpha$, 
 and all equipotentials are closed lines around either a
maximum or a minimum, except those for $V_{\mathrm{eff}} ({\bf r})=0$, which
are straight lines.
We find that the angular velocity of the GC
drift along the equipotentials is given only by the parameter $\Omega$ defined
in Eq.~(\ref{Omega-har}), and geometrical factors. As a consequence, the
suppression of the GC contribution to the conductivity can be
expressed by a function $ \Phi(\Omega \tau)$, and instead of
Eq.~(\ref{dsig-1d}) 
we obtain 
\begin{equation}   \label{diften-fvonX}
\sigma^{\rm 2D}_{xx} = \sigma^{\rm 2D}_{yy}= \frac{e^2m}{\pi \hbar^2}
\frac{\Omega^2\tau}{2 K^2} \, \Phi(\Omega \tau)
\, ,
\end{equation}
and $\sigma^{\rm 2D}_{xy} = \sigma^{\rm 2D}_{yx}= 0$.
The actual calculation of  $ \Phi(\Omega \tau)$ is sketched
 in Appendix~\ref{app1}. The numerical results are  plotted  as
diamonds in  Fig.~\ref{fig:fvonXiso}, together with some analytical
approximations, which are obtained from the asymptotic behavior of the correct
 result for small and large values of $\Omega \tau$ (see Appendix~\ref{app2}). 
 Apparently the three-parameter interpolation formula
\begin{equation}   \label{tre-par-int}
\mathrm{\Phi_3}(\Omega \tau)\!=\![1\!+\!0.25(\Omega \tau)^2]/[1\!+\!0.75(\Omega
\tau)^2 \!+\!0.076\,(\Omega \tau)^4]
\end{equation}
 provides a very good fit to the correct numerical result for all values of
$\Omega \tau$. Note that $\Phi(\Omega \tau) \rightarrow 1$ for $\Omega \tau
\rightarrow 0$, as we expect for the weak-modulation limit.

We want to emphasize that, for the square-symmetric harmonic cosine modulation
the suppression of the GC induced contribution to the
conductivity is
described by the single parameter $\Omega \tau$ which,
 according to Eq.~(\ref{Omega-har}), itself depends on
modulation strength and period, and on the cyclotron radius $R=v_F/\omega_0$.

\begin{figure}[h]  \centering
\includegraphics[width=\linewidth]{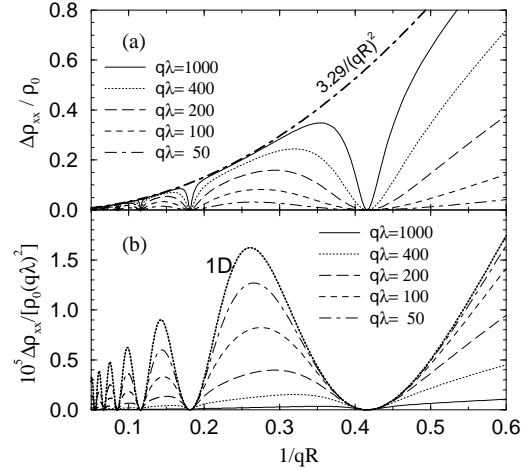}
\caption{\small (a) GC drift contribution to the conductivity
  versus  magnetic field in units $1/qR$ for
 electric modulation $V(x,y)=0.02E_F [\cos qx + \cos qy]$ and several values
  of the mean free path between $q\lambda=50$ (lower dash-dotted) and 1000
 (upper solid 
 line). (b) Same resistance data divided by
  $(q\lambda)^2$, compared with result for the one-dimensional modulation 
$V(x)=0.02E_F \cos qx$ (thick line) which is independent of $q\lambda$.
\label{fig:sqrxx}}
\end{figure}
As an  instructive example, we plot in Fig.~\ref{fig:sqrxx}(a),
 under the assumption $\omega_0\tau \gg 1$, the GC contribution 
 $\Delta \rho_{xx} /\rho_0 \approx (\omega_0\tau)^2 \Delta \sigma_{yy}
 /\sigma_0$ for the electric modulation $V(x,y)=0.02E_F [\cos qx + \cos qy]$
 and several values of the mean free path $\lambda= v_F \tau$
 ($\sigma_0=1/\rho_0=e^2n_{\rm  el}\tau/m$). 
Away from the flat band conditions, given here by the zeroes of the Bessel
function $J_0$, $\Omega \tau$ becomes large with large mean free path, and 
$\Delta \rho_{xx} /\rho_0 \approx (\Omega \tau)^2 \Phi(\Omega \tau)/(qR)^2$
approaches the limiting curve $3.29/(qR)^2$, which is also indicated in 
Fig.~\ref{fig:sqrxx}(a). We see that, as compared with the Drude resistance
$\rho_0$ of
the homogeneous system, the modulation induced correction to the resistance
increases with increasing mean free path and finally saturates.
Since the shape of the resistivity curves in Fig.~\ref{fig:sqrxx} depends only
on the parameter $\Omega \tau$ and $\Omega$ is proportional to the modulation
strength,  variation of the modulation strength  leads to a set of
curves similar to that shown in Fig.~\ref{fig:sqrxx}(a) for the variation of
$(q\lambda)^2$. With increasing modulation strength the curves will 
saturate and approach the same limiting curve, indicated as thick
dash-dotted line in Fig.~\ref{fig:sqrxx}(a).

On the other hand, if we compare this modulation
correction calculated for the square symmetric case with that obtained for the
corresponding 1D modulation, we find, with increasing mean free path,  an
increasingly strong suppression of the WO. This becomes evident from
Fig.~\ref{fig:sqrxx}(b), 
where we have divided $\Delta \rho_{xx} /\rho_0$ by $(q\lambda)^2$, since with
this normalization the 1D result becomes independent of $q\lambda$.
In this plot the suppression of the resistivity maxima
becomes stronger with increasing mean free path and increasing modulation
strength. 

\begin{figure}[h]  \centering
\includegraphics[width=\linewidth]{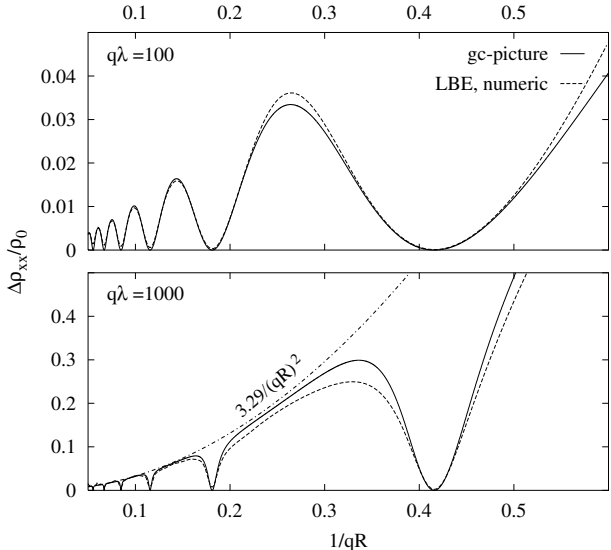}
\caption{\small Comparision of resistivities calculated for 
 electric modulation $V(x,y)=0.02E_F [\cos qx + \cos qy]$
within the GC picture (solid lines) and from numerical solution of
the linearized Boltzmann equation (dotted lines), for two values of the mean
free path.
\label{fig:zwer}}
\end{figure}
To show that the GC picture yields reasonable results, we compare
in Fig.~\ref{fig:zwer} calculations based on Eqs.~(\ref{diften-fvonX}) and 
(\ref{tre-par-int}) with results obtained numerically from the linearized
Boltzmann equation. The
Boltzmann equation was solved by Fourier expansion  of the distribution
function with respect to the
periodic position variables $x$ and $y$ and the angle $\varphi$ in velocity
space, similar to the procedure described in Ref. \cite{Menne98:1707} for a 1D
LSL. To obtain the curve for $q\lambda = 1000$ with sufficient accuracy, 
about $40\, 000$ Fourier coefficients had to be included, and the calculation,
using an optimized parallel code, took about 6
hours on a CRAY-T3E with 128 nodes. The comparison shows that the 
GC approach with the approximation (\ref{tre-par-int}), which
requires only negligible numerics, yields surprisingly  good results for weak
modulations.
The agreement will become poorer for stronger modulation and for much smaller
$q\lambda$. Then, with decreasing $B_0$,  the maxima of the $\Delta \rho_{xx}$
oscillations in the GC approach will still extrapolate to zero, 
whereas the correct calculation yields damped oscillations around a nearly
constant, finite $\Delta \rho_{xx}$ value. But this difference occurs also for
1D LSLs and is well understood.\cite{Menne98:1707}

We conclude that the GC approach yields  reasonable results  for not too
strong modulations (and not too small $q\lambda$ values),  and we will use it
as a versatile approach to discuss interesting situations of lower symmetry. 

Pure magnetic modulations lead to similar results as pure electric
modulations, of course with modifications due to the differences between the
Bessel functions $J_1$ and $J_0$, notably a phase shift.
 Interesting new
situations occur for mixed electric and magnetic modulations, which can be
achieved experimentally, e.g.,  by bringing a rectangular pattern of
ferromagnetic islands on the surface of the sample. \cite{Ye96:1613} 
Superpositions of harmonic electric and magnetic modulations, eventually with
a phase shift, can easily be evaluated using Eq.~(\ref{tre-par-int}), provided 
the effective potential according to Eq.~(\ref{veff-coef}) has square symmetry.

\subsection{Harmonic LSL with rectangular symmetry}
A LSL with exact square-symmetry is an idealized limiting case and hard to
realize experimentally. Therefore, we now consider the more general case of a
rectangular LSL, which allows to interpolate between 1D and square-symmetric
2D modulations, and to approach both limiting cases. To keep the discussion
simple, we restrict it on  harmonic electric and magnetic modulations in both
directions, so that the effective potential  is of the form
\begin{equation} \label{veff-add-cos}
V_{\mathrm{eff}}({\mathbf{r}})=\tilde{V}_x \cos (K_xx+ \alpha_x) +\tilde{V}_y
\cos ( K_yy +\alpha_y) \, ,
\end{equation}
where the ratios of
amplitudes and phases may depend on the amplitudes and relative
phases between the electric and magnetic modulations in $x$ and in $y$
direction, and, in contrast to  Eq.~(\ref{cossum_squ}),
on the average magnetic field $B_0$.

Besides its simplicity, this model is important for the physical reason,
that higher modulation harmonics  decrease
exponentially with the distance of the 2D EG from the surface if the
modulation is produced by some type of surface structuring. Thus, if this
distance is large enough, it will be sufficient to consider only the basic
cosine modulation.

\subsubsection{Numerical and analytical results}
For a given modulation, the ratio $\tilde{V}_y /\tilde{V}_x $ in
Eq.~(\ref{veff-add-cos}) may change  magnitude and sign as a function of
$B_0$. This can lead to interesting switching effects which we will discuss
below. 
For the calculation of the conductivity components (Appendix~\ref{app1}), we
assume however always $0\! \leq \!\kappa \!= \! \tilde{V}_y 
/\tilde{V}_x \! \leq \! 1$, which may eventually require 
an interchanging of $x$ and $y$ in the final results. Then, with $V_{\rm
  cha}=\tilde{V}_x$ in 
Eqs.~(\ref{wxieta}) and (\ref{Kap-omega}), and with a suitable choice of the
origin, the dimensionless potential 
(\ref{wxieta}) becomes
\begin{equation}  \label{cossum_kappa}
w(\xi,\eta)=\cos \xi + \kappa \cos \eta  \, , \quad 0 \leq \kappa \leq 1 \,.
\end {equation}
For $\kappa =0$
we have the 1D modulation in $x$ direction, and the equipotentials are
straight lines in $y$ direction. For $\kappa=1$ we have the square-symmetric
case where all equipotentials are closed lines.
These cases have been considered above. For $0<\kappa <1$, there exist closed
equipotentials with $w(\xi,\eta)=\epsilon$ around maxima in the energy
interval $1-\kappa < \epsilon \leq  1+ \kappa$, closed equipotentials around
minima in the interval $-(1+ \kappa) \leq \epsilon < -(1-\kappa)$, and open
equipotentials in $y$ direction for $-(1-\kappa) \leq \epsilon \leq 1-\kappa$.
We can show that, in the limit of large mean free path ($\tau \rightarrow
\infty$), the GC contribution $\Delta \sigma_{xx}$ comes only from
closed orbits, and shows a suppression similar to that obtained in the
square-symmetric case. The contributions to $\Delta \sigma_{yy}$, on the other
hand, come from both closed and open equipotentials. The latter lead to an
increase with increasing $\tau$, similar to the 1D case.

 The off-diagonal components $\Delta \sigma_{xy}=\Delta \sigma_{yx}=0$ can be
 shown to vanish from symmetry reasons.
The analytical considerations of Appendix~\ref{app1} show that the diagonal
components can be written as
\begin{equation}   \label{diften-fmuvonX}  
\Delta \sigma_{\mu \mu} = \frac{e^2 m}{\pi \hbar^2}\, \frac{\Omega^2\tau}{2
  K_{\mu}^2}\, \Phi_{\mu \mu} (\Omega\tau,\kappa) \, ,
\end{equation}
with $\Omega = K_x K_y \tilde{V}_x/(m \omega_0)$, so that the (suppression)
effect of the 2D modulation now is described by two parameters, $ \Omega\tau$
and $\kappa$. We have numerically calculated the functions $ \Phi_ {\mu \mu}
(\Omega\tau,\kappa)$, which of course satisfy the consistency relations 
$\Phi_{xx}(\Omega\tau,1)\equiv \Phi_{yy}(\Omega\tau,1)\equiv \Phi
(\Omega\tau)$. Since it is rather time-consuming to calculate the successive
fourfold integrals with sufficient acccuracy for each specific example anew,
we tried to fit the $ \Phi_ {\mu \mu}(\Omega\tau,\kappa)$ by simple analytic
expressions.  
We found that the numerical results for  $ \Phi_{xx}(\Omega\tau,\kappa)$ 
are very well (with an error of less than 1 per cent) approximated by
$\kappa^2 \Phi(\Omega\tau)$, so that a good approximation is
\begin{figure}[h]  \centering
\noindent  
\includegraphics[width=\linewidth]{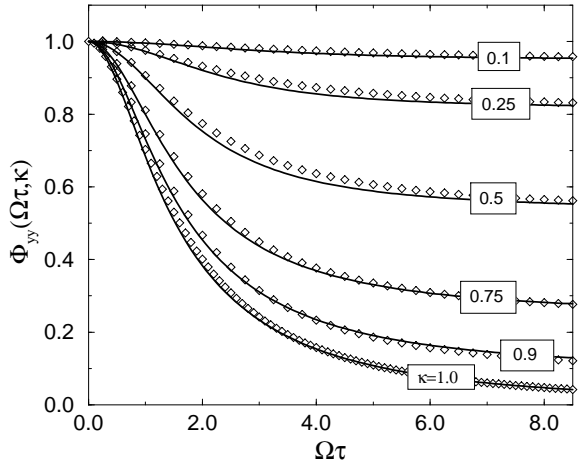}
\caption{\small Numerical result  for
  $\Phi_{yy}(\Omega\tau,\kappa)$  (diamonds) for several values of the
anisotropy parameter $\kappa
  = \tilde{V}_y/ \tilde{V}_x$. The solid lines are for the approximation
    $\Phi_{yy}^{(3)}$, Eq.~(\protect\ref{fiyy_3}). 
\label{fig:syybest}}
\end{figure}
\begin{equation}   \label{fixx-appr}
\Phi_{xx}(\Omega\tau,\kappa) \approx \kappa^2 \Phi(\Omega\tau)
\approx \kappa^2 \Phi_3(\Omega\tau) \, ,
\end{equation}
with $\Phi_3(\Omega\tau)$ defined by Eq.~(\ref{tre-par-int}).
Numerical results for $ \Phi_{yy}(\Omega\tau,\kappa)$ are shown as diamonds
 in Fig.~\ref{fig:syybest}. Apparently, for $\kappa <1$ they
approach a finite limit for $ \Omega\tau \rightarrow \infty$. This limit 
$\Phi_{yy}(\infty,\kappa)$ is easily calculated numerically and well
approximated by 
\begin{equation}   \label{two-pwr-interp}
\Phi_2(\kappa)=1-1.645\, \kappa^{3/2} +0.645\, \kappa^{5/2} \, ,
\end{equation}
 see Appendix~\ref{app2}. Incorporating this into an
 interpolation formula  that reduces for $\kappa =1$ to the
previous fit (\ref{tre-par-int}), we obtained
\begin{equation}   \label{fiyy_3}
\Phi_{yy}^{(3)}(\Omega\tau,\kappa)=\Phi_2(\kappa)+\,
\frac{[1-\Phi_2(\kappa)] [1+\alpha_\kappa\, (\Omega\tau)^2]}
{1+(\alpha_\kappa+\beta_\kappa)\, (\Omega\tau)^2+\gamma_\kappa \,
  (\Omega\tau)^4}\, , 
\end{equation}
with $\alpha_\kappa =0.25 \sin^2(\pi\kappa/2)$, $ \beta_\kappa=
0.5 \kappa^2/[1-\Phi_2(\kappa)]$, and
$\gamma_\kappa=0.076  \sin^2(\pi\kappa/2)$.
This approximation is indicated by the lines in
Fig.~\ref{fig:syybest} and will in the following be used instead 
 of $ \Phi_{yy}(\Omega\tau,\kappa)$.

\subsubsection{Two examples}

\begin{figure}[h]  \centering
\includegraphics[width=\linewidth]{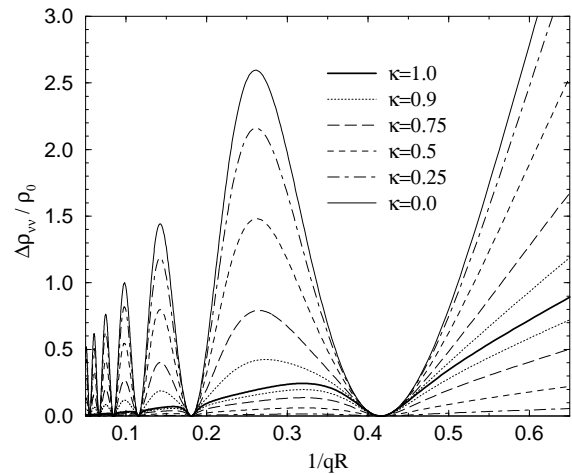}
\caption{\small GC contribution to the resistivities for the
  electric modulation  
$V(x,y)/E_F= 0.02\, [\, \cos (qx)$ $+\kappa \cos(qy)\,]$ and
$q\lambda=400$ . For $\kappa=1$ (thick 
line) $\Delta \rho _{xx}=\Delta \rho _{yy}$. For $\kappa <1$,  the result for
$\Delta \rho _{yy}$ lies below, that for $\Delta \rho _{xx}$ above this thick
line, and both are indicated by the same line style. \label{fig:sxxyyanis}}
\end{figure}

First we consider in Fig.~\ref{fig:sxxyyanis}  a purely electrostatic
modulation on a square lattice, 
$a_x=a_y=2\pi/q$, $V(x,y)=V_x \cos (qx) + V_y \cos(qy)$, so that
$\tilde{V}_{\mu} = V_{\mu} |J_0(qR)|$ and the ratio $\kappa =\tilde{V}_y
/\tilde{V}_x =V_y /V_x$ is independent of $B_0$, and $\Omega =q^2  V_x
|J_0(qR)|/(m \omega_0)$. 
For $\kappa \! <\!1$, there exist open equipotentials only in $y$
direction. With 
decreasing $\kappa$ their number increases, and $\Delta \rho _{xx}$ increases
towards the results for the 1D modulation
($\kappa\!=\!0$). Simultaneously $\Delta \rho _{yy} \propto \kappa^2$
decreases, 
and vanishes in the 1D limit. 
The degree of anisotropy increases with both the modulation amplitude and the
mean free path, since, for $\Omega\tau \gg 1$,
  $\Delta \rho _{yy}/\rho_0\approx 3.29 \kappa^2/(qR)^2$ saturates, while 
$\Delta \rho _{xx}/\rho_0 \sim (\Omega\tau)^2 \Phi_2(\kappa)/(qR)^2$ increases
without limit.

The anisotropy parameter $\kappa$ is only a constant independent of $B_0$ if
we have either a pure electric or a pure magnetic modulation on a square
lattice, i.e., with the same period in $x$ and $y$ direction. In all other
situations, the Bessel functions in Eq.~(\ref{veff-coef}) lead to a
$B_0$-dependent $\kappa$. In such cases we use the following convention to
express the relevant parameters $\Omega$ and $\kappa$ in terms of the original
parameters specifying the modulation.

We measure energies in units of $E_F=mv_F^2/2$ and  the average magnetic field
in dimensionless units 
$1/(qR)$, where $R=v_F/\omega_0$ is the cyclotron radius and $q=\sqrt{K_x
  K_y}$. Then, for a suitable choice of the coordinate system, the modulation
may depend on the following seven parameters: 
(1) the ratio of the lattice constants $ a_y/a_x = K_x/K_y$, (2) the
amplitudes $\varepsilon_{\nu} = V_{\nu}/E_F$ of the electric cosine potential
$V(x,y)=E_F [\varepsilon_x \cos (K_xx)+\varepsilon_y \cos (K_yy)]$, (3)
 the amplitudes   $\mu_{\nu} = 2 \omega_{\nu}/(K_{\nu} v_F)$ and (4) the
 relative 
 phases  $\alpha_{\nu}$ of the effective magnetic modulation potential 
$E_F [\mu_x \cos (K_x x+ \alpha_x) + \mu_y \cos(K_y y+\alpha_y)]$.
For each value of the average magnetic field, we can calculate from these
{\em seven} model parameters the {\em two} parameters of the effective
potential (\ref{veff-coef}) which are relevant for the conductivity, namely
the absolute values of the complex numbers $\varepsilon_{\nu} J_0(K_{\nu} R)
+\mu_{\nu} J_1(K_{\nu} R) \exp(i \alpha_{\nu})$,
\begin{eqnarray}
\tilde{\varepsilon}_{\nu}=&&\left\{ \left[ \varepsilon_{\nu} J_0(K_{\nu} R) 
+\mu_{\nu} J_1(K_{\nu} R)\cos \alpha_{\nu}\right]^2 \right.\nonumber \\
&& \left. \hspace*{2.1cm}
+\left[\mu_{\nu} J_1(K_{\nu} R)\sin \alpha_{\nu} \right]^2\right\} ^{1/2}
 \label{veffnu}
\end{eqnarray}
for $\nu=x,\,y$. The  phases of these complex numbers can be 
compensated by a suitable shift of the coordinate system and have no effect on
the conductivity. In the following we use these two parameters in the form
${\varepsilon}_{\mathrm{max}}=
\max[\tilde{\varepsilon}_x,\tilde{\varepsilon}_y]$
and $\kappa=\min[\tilde{\varepsilon}_x,\tilde{\varepsilon}_y]
/\max[\tilde{\varepsilon}_x,\tilde{\varepsilon}_y]$.
Taking the characteristic energy in Eq.~(\ref{Kap-omega}) as 
$V_{\mathrm{cha}}={\varepsilon}_{\mathrm{max}} E_F$, we obtain
$\Omega=\omega_0 {\varepsilon}_{\mathrm{max}}(qR)^2/2$.

 To characterize the system
completely, we have to specify the mean free path $\lambda=v_F \tau$, which we
write in the dimensionless form $q\lambda$, so that
$\omega_0\tau=\lambda/R$. Finally we obtain for the GC drift
contribution to the conductivity tensor
 \begin{equation} \label{sigma-scaled}
\frac{\Delta \sigma _{\mu \mu}}{\sigma_0} =\frac{q^2}{4K_{\mu}^2}\,
(qR{\varepsilon}_{\mathrm{max}})^2 \, \Phi_{\tilde{\mu}\tilde{\mu}}
\Big(\frac{1}{2}q^2\lambda R {\varepsilon}_{\mathrm{max}},\kappa\Big) \, ,
\end{equation}
with $\tilde{x}=y$ and $\tilde{y}=x$ if
${\varepsilon}_{\mathrm{max}}=\tilde{\varepsilon}_y$, and with
$\tilde{x}=x$ and $\tilde{y}=y$ if
${\varepsilon}_{\mathrm{max}}=\tilde{\varepsilon}_x$. 
Since in the regime of commensurability oscillations $\omega_0\tau \gg 1$, the
GC correction to the resistivity tensor is
 $\Delta \rho _{\mu \mu}/\rho_0= (\omega_0\tau)^2 \Delta \sigma _{\bar \mu
  \bar \mu}/\sigma_0$, with $\bar x=y$ and $\bar y=x$.

\begin{figure}[h]  \centering
\includegraphics[width=\linewidth]{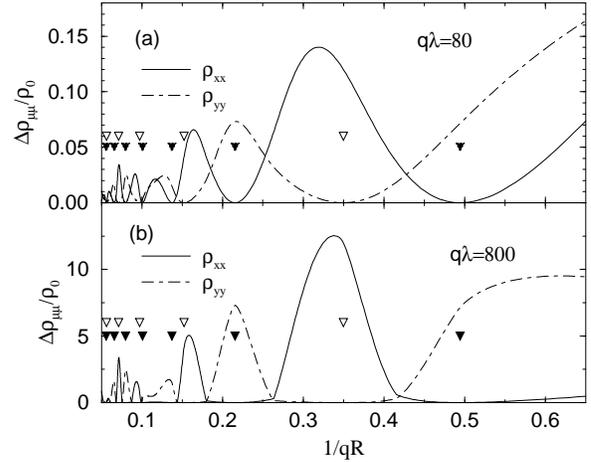}
\caption{\small GC contribution to the resistivities for
  modulation  
$V(x,y)/E_F= 0.02\, [\cos (K_xx)+\cos(K_yy)]$, with $K_x/K_y=\sqrt{2}$,
versus magnetic field in units of $1/qR$, with $q=\sqrt{K_xK_y}$; (a) for
$q\lambda =80$, (b) for $q\lambda =800$. The ``flat-band'' conditions
$J_0(K_xR)=0$ and $J_0(K_yR)=0$ are indicated by filled and open triangles,
respectively. \label{fig:diagitter}} 
\end{figure}

As a  very interesting example we consider  a purely electrostatic
modulation, but now on an rectangular superlattice with  equal modulation
amplitudes $\varepsilon_x=\varepsilon_y$ but
different periods in
both directions, $a_y/a_x=\sqrt{2}$.
The interesting aspect of this model is that now the effective potential
changes its symmetry as a function of the magnetic field strength, since the
arguments of the Bessel functions in Eq.~(\ref{veff-coef}) are different. If
one of the Bessel functions vanishes, i.e.\ if the ``flat-band''
condition for this direction is satisfied, the effective potential shows a
purely 1D modulation in the other direction. 
When the effective modulation potential in $x$ direction is larger than that
in $y$ direction, there exist open equipotentials in $y$ but not in $x$
direction, and vice versa. Typical results for
the resistivity corrections are shown in Fig.~\ref{fig:diagitter}. For
relatively small mean free path as in Fig.~\ref{fig:diagitter}(a), the
oscillations of the resistivity 
components $\Delta \rho_{\mu \mu}$   look similar to the results one would
expect for the corresponding 1D modulations. At relatively low
magnetic fields, there occurs however a kind of beating effect, manifested in
a non-monotonous decrease of the oscillation amplitude of $\Delta 
\rho_{xx}$ (solid line) with decreasing magnetic field $B_0$. The reason for
this non-monotonous $B_0$-dependence of the maxima is easily understood. The
maxima occur nearly in the middle between adjacent flat-band conditions
$J_0(K_xR)=0$. If for these $B_0$-values
the effective modulation in $y$ direction is large [i.e., if no zero of
$J_0(K_yR)$ is close], the GC motion is essentially
two-dimensional, and the maximum of $\Delta\rho_{xx}$ is suppressed below the
corresponding one of a 1D modulation in $x$ direction. If,
however, the maximum of $\Delta\rho_{xx}$ appears near a zero of $J_0(K_yR)$,
the modulation in $y$ direction is small, and the $\Delta\rho_{xx}$ maximum
assumes a large value close to that of the corresponding 1D modulation in $x$
direction. This explains why the $\Delta\rho_{xx}$ maximum near
$(qR)^{-1}=0.071$ is higher than those near 0.091 and 0.116. 
These anisotropy effects are drastically enhanced for larger mean free path,
see  Fig.~\ref{fig:diagitter}(b). If, for example, at a given $B_0$ value the
effective modulation potential in $y$ direction is smaller than that in $x$
direction, the contribution to $\Delta\rho_{yy}$ ($ \propto\Delta\sigma_{xx}$)
results only from guiding 
centers moving along closed equipotentials, and $\Delta\rho_{yy}/\rho_0$
is bounded by $3.29(\tilde{\varepsilon}_y/\tilde{\varepsilon}_x)^2/(qR)^2
\ll 3.29/(qR)^2$ (cf. Fig.~\ref{fig:sqrxx}). There exist, however,
open equipotentials in $y$ direction which lead to an increase of
$\Delta\rho_{xx} \propto (q\lambda)^2$ with increasing mean free path. The
result is a very effective switching as a function of the magnetic field $B_0$
between regions with large $\Delta\rho_{xx}$ and very small $\Delta\rho_{yy}$
and regions with small $\Delta\rho_{xx}$ and large $\Delta\rho_{yy}$, as is
seen in Fig.~\ref{fig:diagitter}(b).

If one mixes electric and magnetic modulations with different phase shifts in
both directions, one may achieve such  switching effects also on a
square lattice, $K_x=K_y$.

\section{Summary}

We have evaluated the modulation correction to the magneto-resistivity tensor
of 2D EGs in  LSLs of rectangular symmetry within the 
GC picture. We have emphasized that this classical approach can be
useful only within a restricted regime of sufficiently weak modulations and
sufficiently strong (average) magnetic fields, where the electron motion may
be  approximated  as  a rapid cyclotron motion around
slowly drifting GCs.
Within this regime, in which a 2D EG with a 1D LSL exhibits  regular
commensurability oscillations  (WO), we have
investigated the effects of the model parameters (modulation strengths,
anisotropy, phase shifts) and the  mean free path on the amplitudes of the WO.
For harmonic electro- and 
magnetostatic modulations we have obtained essentially analytical results. 

The fact that the GCs move approximately along the equipotentials
of a magnetic-field dependent effective  potential, with a velocity 
essentially proportional to the strength of this potential, leads to an
interesting dependence of the WO amplitudes on these model parameters. In
contrast to a 1D LSL, which has only extended straight-line equipotentials, a
2D LSL has also closed equipotentials around the extrema of the effective
potential.
The difference between closed and extended equipotentials becomes important in
the limit of a large mean free path, since the magnetoresistance is sensitive
to the mean velocity of the GC motion between two successive
scattering events. If the scattering time is sufficiently large, the
GC velocity along closed equipotentials averages to zero, and then
these equipotentials do not contribute to the magnetoresistance, whereas
the contribution of extended equipotentials becomes very large.

 This leads to strongly anisotropic resistivities, if the effective potential
 has 
 rectangular but not square symmetry, and to very interesting magnetic-field
 dependent switching effects if the symmetry of the effective potential
 changes as a function of the average magnetic field.

For the 2D EG with a weak square-symmetric modulation we find with increasing
 mean free path  an increasing suppression of the WO amplitudes below
 those obtained for the corresponding 1D modulation. This result provides a
 classical explanation of the suppression of the band-conductivity observed in
 early experiments on holographically modulated high-mobility samples,
 \cite{Weiss90:88} which previously has been explained with quantum arguments
 based on the subband splitting of the Hofstadter energy spectrum.
 \cite{Gerhardts91:5192,Pfannkuche92:12606}
 For fixed mean free path, our
 (basically analytical) result reduces in the limit of very weak modulations
 to the predictions of Ref.\cite{Gerhardts96:11064}, without noticeable
 suppression of the WO. For realistic values of modulation strength and mean
 free path, our present results yield, however, a strong  suppression.
We want to point out that our present classical explanation of the WO
 suppression and the previous quantum one are not contradictory. Both need for
 the explanation of an effective suppression a sufficiently strong modulation
 and   large mean free path (i.e.\ weak disorder). 

Qualitatively our result is also in agreement with the recent prediction 
of the suppression of WO by  Grant {\em et al.} \cite{Grant00:13127},   
which applies to the case of intermediate mean free path and strong 2D
 modulation. From our investigation of trajectories we expect, however, that
 for this strong modulation the regime of small and intermediate values of the
 average magnetic field is dominated by chaotic motion, so that the
 GC picture cannot be expected to yield quantitatively correct
 results.

Finally we want to comment on the fact that the Chambers formula
(\ref{diff-tens-gc}) contains a scattering time, which describes isotropic
impurity scattering, whereas calculations for the 2D EG with a 1D LSL based on
Boltzmann's equation  have
revealed that predominantly small-angle impurity scattering has to be
considered for a quantitative understanding of the WO amplitudes.  
We did not try to go beyond the simple relaxation time approximation in the
GC picture, since (i) on the level of Boltzmann's equation, where
we know how to describe anisotropic scattering, we cannot separate the
GC from the cyclotron motion, and (ii) small-angle scattering of
an electron between locally nearby trajectories may include large changes of
the corresponding GCs, and we do not want to introduce unjustified
assumptions on scattering between GCs. In view of the general
limitations of the GC picture, we rather want to consider the
relaxation time $\tau$ as a phenomenological parameter, which may be chosen to
fit experiments qualitatively. We think, however, that $\tau$ should be
considered as the total scattering time, which in the case of strongly
anisotropic impurity scattering is much shorter than the transport or momentum
relaxation time.

\acknowledgments
We are grateful to J.~H.~Davies and A.~R.~Long for a stimulating discussion
on the guiding-center picture and to C. Albrecht for useful suggestions
concerning the manuscript. This work was supported by BMBF Grant
No. 01BM919/5.  

\appendix
\section{Correlation integrals along equipotentials  } \label{app1}
To study the effect of closed equipotentials, we
 assume that the effective potential $w(\xi,\eta)$ has either isolated
maxima or isolated minima, or both. For instance, the model
(\ref{cossum_kappa}) has, for arbitrary integers $m$ and $n$, isolated maxima
 at $(\xi , \eta)=2\pi(m,n)$ and isolated minima at  $(\xi ,
 \eta)=(2m+1,2n+1)\pi$, and all equipotentials $w(\xi,\eta)=\epsilon$ for
 $|\epsilon |>1-\kappa$ are closed.

We assume that closed  equipotentials around a maximum (minimum), which we
take as 
origin, exist in the energy interval $\epsilon_{\mathrm{max}} \geq \epsilon >
 \epsilon_{\mathrm{sup}}$ ($\epsilon_{\rm min} \leq \epsilon < \epsilon_{\rm
   inf}$). In terms of polar coordinates,
\begin{equation} \label{polar}
\xi = \rho \cos \varphi, \quad \eta= \rho \sin \varphi, \quad -\pi <  \varphi
\leq \pi  \, ,
\end{equation}
the equipotential with energy $\epsilon$  is described by the equation 
 $\rho = \rho_{\epsilon} (\varphi)$, which maps $ \varphi$ onto the solution
$ \rho$ of $w( \rho \cos \varphi,\, \rho \sin \varphi)=\epsilon$ for fixed
 $\epsilon$ and $ \varphi$.
Along the equipotential with energy $\epsilon$ the equations
 (\ref{xi-eta-dot}) reduce to
\begin{equation} \label{dfidt-clmax}
d\varphi /dt =\pm \Omega /\mathcal{J}_{\epsilon} (\varphi) \, ,
\end {equation}
where the upper (lower) sign stands for orbits around a maximum (minimum), and 
\begin{equation} \label{jacobian}
{\mathcal{J}}_{\epsilon} (\varphi) = \left| \frac{\rho_{\epsilon}
  (\varphi)}{\cos 
  \varphi   \, w_{\xi} + \sin \varphi \, w_{\eta}}\right|  \, .
\end {equation}
We can use  Eq.~(\ref{dfidt-clmax}) to substitute in
Eq.~(\ref{diff-tens-gc}) the integration variable $t$ by $\varphi$. Writing
the initial position on an equipotential as ${ \mathbf{r}}(0)= \rho_{\epsilon}
(\varphi_0)( \cos \varphi_0 ,  \sin \varphi_0)$, we get
$t= \pm \int_{ \varphi_0}^\varphi d\varphi'
{\mathcal{J}}_{\epsilon}(\varphi')/\Omega $.
 With $2\theta_{\epsilon} =
\int_{-\pi}^{\pi} d\varphi\,  {\mathcal{J}}_{\epsilon} (\varphi)$ one obtains
\begin{equation}
\int _0^\infty\!\! dt \, e^{-t/\tau} w_\mu (\varphi(t); \epsilon) =
\frac {\tau X \, W_\mu^{\mp}( \varphi_0, \varphi_0  \pm 2\pi; \epsilon ) }{1-
  e^{-2X\theta_{\epsilon}}} , 
\end{equation}
where  $X=1/(\tau\Omega)$ and 
\begin{equation} \label{w-hilf}
W_\mu^{\pm}( \varphi_0, \varphi_1;\epsilon ) \!  = \! \mp \!
 \int_{ \varphi_0}^{ \varphi_1} \!\!
d\varphi\, {\mathcal{J}}_{\epsilon} (\varphi)\,  w_\mu (\varphi; \epsilon) \,
 e^{ \pm X \!  \int_{ \varphi_0}^\varphi \! d\varphi' 
 {\mathcal{J}}_{\epsilon} (\varphi')} \, . 
\end{equation}

To evaluate the average over initial values in Eq.~(\ref{diff-tens-gc}), we
first integrate along the equipotentials with fixed energy $\epsilon$ and 
then over $\epsilon$. It turns out that the Jacobian of the 
transformation from polar coordinates  $\rho, \, \varphi$ to the energy-angle
coordinates $\epsilon, \, \varphi$ is just given by Eq.~(\ref{jacobian}),
$d\rho\, \rho \, d \varphi = d\epsilon \, d \varphi \, {\mathcal{J}}_{\epsilon}
(\varphi)$.
If the effective potential $w(\xi,\eta)$ is an even function of
both arguments, we have $w_{\mu}(\varphi+\pi; \epsilon )= - w_{\mu}(\varphi;
\epsilon )$ and $ {\mathcal{J}}_{\epsilon} (\varphi
+\pi)={\mathcal{J}}_{\epsilon} 
(\varphi)$, and all integrals over intervals of length $2\pi$ can be reduced
to integrals over intervals of length $\pi$, and we obtain

\begin{eqnarray}
D^{{\mathrm{cl}},\pm } _{\mu \nu} =&&  \frac{\sigma_{\mu} \sigma_{\nu}}{K_{\mu}
  K_{\nu}} \frac{\Omega^2\tau}{(2\pi)^2}  
\int_{ a_{\pm}}^{b_{\pm}}  d \epsilon
\int_0^\pi d \varphi_0\,{\mathcal{J}}_{\epsilon} (\varphi _0) \nonumber \\
&&\hspace*{0.5cm} \times  w_{\bar{\nu}} (\varphi _0; \epsilon) 
   \frac{2 X \,
W_{\bar{\mu}}^{\mp}( \varphi _0, \varphi _0 \pm \pi; \epsilon )}{1+e^{-X
  \theta_{\epsilon}}}   \,  \label{diften-clmax}
\end{eqnarray}
with the upper sign and $a_+=\epsilon_{\rm sup}$, $b_+=\epsilon_{\rm max}$
for equipotentials around a maximum and the lower sign and  $a_-
=\epsilon_{\rm min}$, $b_-= \epsilon_{\rm inf}$ for those around a minimum.

 Open equipotentials of a periodic potential with
rectangular symmetry, by definition,
 connect one point on a boundary of the unit cell with the
equivalent point on the opposite boundary. Since equipotential lines cannot
cross each other, open equipotentials can exist either in $x$ direction or in
$y$ direction, but not in both. Let us assume that in the energy interval  $
\epsilon_{\mathrm{inf}} \leq \epsilon \leq \epsilon_{\mathrm{sup}}$ open
equipotentials in $y$ direction exist. We may describe them in polar
coordinates choosing the origin in a maximum, so that we can use the formalism
developed above. Let the equipotential with energy $\epsilon$ hit
the upper boundary of the unit cell at $\eta = \pi=
\rho_{\epsilon}(\varphi_{\epsilon}) \sin \varphi_{\epsilon}$ for $\pi /4 \leq
\varphi_{\epsilon} < \pi/2$. Assuming that $w(\xi,\eta)$ is even with respect
to both arguments, we can show that the GC motion in negative $y$
direction with initial conditions $\pi -\varphi_{\epsilon} \leq \varphi_0 \leq 
\pi + \varphi_{\epsilon}$ yields the same contribution to the diffusion tensor
as those moving in the positive $y$ direction with initial conditions
$ -\varphi_{\epsilon} \leq \varphi_0 \leq  \varphi_{\epsilon}$, and we consider
here only the latter.

For the time integration, we divide the infinite time interval into an initial
one of duration $t_{\epsilon}= \Omega^{-1} \int _{\varphi_0}^{ \varphi_{\epsilon}}
\! d \varphi \, {\mathcal{J}}_{\epsilon} (\varphi)$ and subsequent intervals of
duration $T_{\epsilon}= \Omega^{-1} \int _{- \varphi_{\epsilon}}^{
  \varphi_{\epsilon}} \! d \varphi \, {\mathcal{J}}_{\epsilon} (\varphi)$,  
which is the time a GC needs to traverse a unit cell on the
equipotential of energy $\epsilon$. 
Using the definition (\ref{w-hilf}), we obtain for the contribution 
of open orbits to the diffusion tensor
\begin{eqnarray}
D^{\mathrm{open}}_{\mu \nu}&& =
 \frac{\sigma_{\mu} \sigma_{\nu}}{ K_{\mu} K_{\nu}}\,
 \frac{2X\Omega^2\tau }{(2\pi)^2} 
 \! \int_{ \epsilon_{\mathrm{inf}}}^{\epsilon_{\mathrm{sup}}}\!\!  d \epsilon
\! \int_{- \varphi_{\epsilon}}^{  \varphi_{\epsilon}}\!\! d \varphi_0
 \,{\mathcal{J}}_{\epsilon} (\varphi_0) w_{\bar{\nu}} (\varphi_0; \epsilon)
 \nonumber \\ 
&&\hspace*{-0.2cm} \times 
\left\{ 
W_{\bar{\mu}}^-( \varphi_0, \varphi_{\epsilon}; \epsilon )\!
+\! \frac{e^{-t_{\epsilon}/\tau}\,
W_{\bar{\mu}}^- (- \varphi_{\epsilon}, \varphi_{\epsilon}; \epsilon )}{1-
e^{-T_{\epsilon}/\tau}} \right\} .  \label{diff-open}
\end{eqnarray}
To this we have to add the contribution of closed orbits according to Eq.~(\ref{diften-clmax}), $D^{\rm gc}_{\mu
  \nu}= D^{\mathrm{open}}_{\mu \nu} +D^{\mathrm{cl},+}_{\mu \nu}+
  D^{\mathrm{cl},-}_{\mu \nu}$. The result can be written in the form of Eq.~(\ref{diften-fmuvonX}).

\section{Analytic and asymptotic results} \label{app2}
We present explicit results for the additive cosine model
(\ref{cossum_kappa}) of the effective potential, with  $0
\leq \kappa \leq 1$.  The partial derivatives  are then
$w_{\xi}=-\sin \xi$ and $w_{\eta}= -\kappa \sin \eta$. 

We consider first the symmetric case  $\kappa =1$ which, according to
Eq.~(\ref{veff-coef}),  can hold for all
values of the magnetic field only if the original modulation has square
symmetry with equal lattice constants $a_x=a_y=a=2\pi/K$ in both directions,
so that 
$\Omega=(2\pi/a)^2 \tilde{V}_x/(m \omega_0)$.

Things become especially simple  close to the maximum at the origin,
where $w_{\xi} \approx -\xi$ and  $w_{\eta} 
\approx -  \eta$. Then the
equipotentials become circles with radii $\rho_{\epsilon}=4-2\epsilon$
independent of $\varphi$, and the Jacobian (\ref{jacobian}) reduces to
${\mathcal{J}}_{\epsilon} =1$. The angular velocity $d\varphi /dt=\Omega$
becomes  
constant along the equipotentials, and independent of $\epsilon$. Thus, the
GC motion in this approximation is very similar to the simple
cyclotron motion, however with the circular frequency $\Omega$ instead of the
cyclotron frequency $\omega_0$. As a consequence, all integrals in
Eq.~(\ref{diften-clmax}) can easily be evaluated analytically, with the result
\begin{equation} \label{diften-toy}
D_{xx}^{\mathrm{cl,+}}(\epsilon_{\mathrm{qu}})
 = \Big(\frac{a}{2\pi}\Big)^2\, \frac{\Omega^2
  \tau}{(2 \pi)^2} \, \frac{\pi   (2-\epsilon_{\mathrm{qu}})^2}{ 1+(\Omega
 \tau)^2}   \, , 
\end{equation}
where $\epsilon_{\mathrm{qu}}$ is the energy above which the quadratic
approximation is valid, and
$D_{yy}^{\mathrm{cl,+}}= D_{xx}^{\mathrm{cl,+}}$, and 
$D_{yx}^{\mathrm{cl,+}}=-D_{xy}^{\mathrm{cl,+}}= \Omega \tau \,
D_{xx}^{\mathrm{+}}$. 
Thus, for $\Omega \tau \gg 1$, the  motion  of the GCs along closed
 equipotentials leads to a suppression  
 $\propto (\Omega \tau)^{-2}$. For a suitable choice of
 $\epsilon_{\mathrm{qu}}$ ($=2-\sqrt{\pi} =0.228$) and a corresponding
 treatment of $D_{\mu \nu}^{\mathrm{cl,-}}$, one obtains the result indicated
 in Fig.~\ref{fig:fvonXiso} by the long-dashed line.

Going  beyond this simple quadratic approximation, we obtain qualitatively
similar results. All equipotentials
with energy $\epsilon >0\, (<0)$ are closed lines around a maximum
(minimum). As $|\epsilon |$ becomes small, the angular velocity varies along 
the orbits and becomes very small near the saddle points [$(\xi,\eta)=(0,
\pi)$ and equivalent], where the Jacobian ${\mathcal{J}}_{\epsilon}(\varphi)$
diverges.  
 Only the equipotentials exactly at 
$\epsilon=0$ are open trajectories (straight lines), but they yield 
vanishing contribution to the diffusion tensor. 
Exploiting the symmetry, we can show   that $D_{\mu \nu}^{\mathrm{cl,+}}$ and
$D_{\mu \nu}^{\mathrm{cl,-}}$ of Eqs.~(\ref{diften-clmax})  yield identical
contributions to the diagonal components of the diffusion tensor, whereas
their contributions to the off-diagonal components cancel. The result for
the non-vanishing
diagonal components can be written as Eq.~(\ref{diften-fvonX}). The
numerically  calculated $ \Phi(\Omega\tau)$ is plotted in 
Fig.~\ref{fig:fvonXiso}.
In the weak-modulation limit (or ``dirty limit'', $\Omega\tau
\rightarrow 0$) we have  $ \Phi(0)=1$, as expected from
Eqs.~(\ref{diff-t-gc1}) and  (\ref{diff-t-gc-small}). For
$\Omega\tau\rightarrow \infty$, $\Phi(\Omega\tau)$ becomes small.  
We can expand Eq.~(\ref{diften-clmax}) for large
$\Omega\tau$ and show that the term linear in $1/\Omega\tau$ vanishes
identically.  The prefactor of the
leading term  can be 
calculated numerically, and we obtain  $ \Phi(\Omega\tau) \approx 3.29
/(\Omega\tau) ^2$ 
for $\Omega\tau \rightarrow \infty$. This can be used to obtain the
one-parameter 
interpolation  $ \Phi_1(\Omega\tau)=3.29 /(3.29 +(\Omega\tau)^2)$, which 
approximates $ \Phi(\Omega\tau)$ well  for large values of $\Omega\tau$
(see Fig.~\ref{fig:fvonXiso}). An
apparent improvement at small and intermediate $\Omega\tau$ is obtained with
the approximation $ \Phi_3(\Omega\tau)$ defined in Eq.~(\ref{tre-par-int}).

We now turn to the general rectangular symmetry.
For $0 \leq \kappa < 1$ the equipotentials with energies $|\epsilon |\leq
1-\kappa$ are open (in $y$ direction), and degenerate into straight lines 
for $\kappa =0$.

In the ``dirty limit'' $\Omega\tau \rightarrow 0$ the distinction between open
and closed equipotentials is not relevant, since we can
expand the velocity 
$v_{\mu}(t)$ into a Taylor series for small $t$ and perform the integral in
Eq.~(\ref{diff-tens-gc}) term by term. Up to second order in $\Omega\tau$ we
obtain 
\begin{eqnarray}   \label{fixxsmall}
&& \Phi_{xx}(\Omega\tau,\kappa)=\kappa^2\, [1 - (\Omega\tau)^2/2 \, +\dots]\, ,\\
&&   \label{fiyysmall}
 \Phi_{yy}(\Omega\tau,\kappa)=1 - \kappa^2(\Omega\tau)^2/2\, +\dots \, ,
\end{eqnarray}
for all values of $\kappa$.  

\begin{figure}[h]  \centering
\includegraphics[width=\linewidth]{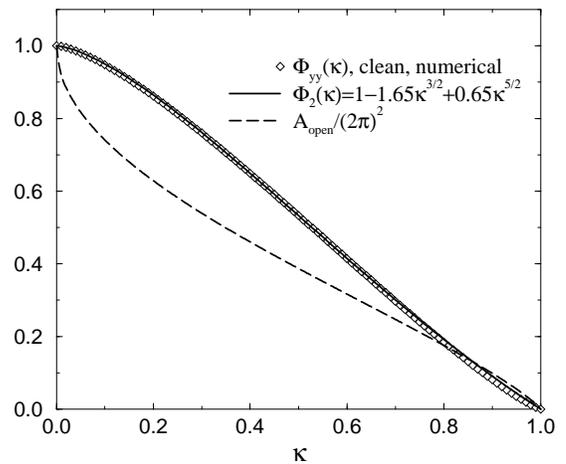}
\caption{\small Numerical result in the clean limit for
  $\Phi_{yy}(\infty,\kappa)$  (diamonds) versus anisotropy parameter $\kappa
  = \tilde{V}_y/ \tilde{V}_x$, together  with analytic approximation
  (solid line). Also shown is the fraction of the unit cell covered by open
  equipotentials   (dashed). 
\label{fig:fvonkappa}} 
\end{figure}

In the ``clean limit'', open equipotentials dominate
$\Phi_{yy}(\Omega\tau,\kappa)$ and introduce a characteristic $\sqrt{\kappa}$
dependence for $\kappa \ll 1$. Already the fraction of the unit cell covered
by open equipotentials,  $A_{\mathrm{open}}/(2\pi)^2=1-
A_{\mathrm{closed}}^{\mathrm{max}}/(2\pi^2)$, which is plotted versus $\kappa$
(as  dashed line) in Fig.~\ref{fig:fvonkappa}, shows such a dependence. To see
that, we calculated the corresponding area
$A_{\mathrm{closed}}^{\mathrm{max}} =\int_{1-\kappa}^{1+ \kappa} d \epsilon  
\int_{-\pi}^{\pi} d\varphi \, {\mathcal{J}}_{\epsilon} (\varphi)$ covered by
closed equipotentials around a maximum (equal to that around a minimum), which
allows the expansion $A_{\mathrm{closed}}^{\mathrm{max}}
=16\sqrt{\kappa} +O(\kappa^{3/2})$. 

The contributions of closed equipotentials to both $ \Phi_{xx}$ and $
\Phi_{yy}$ vanish in the clean limit. The contribution
of open equipotentials to $ \Phi_{yy}(\infty,\kappa)$ is finite while that to 
$ \Phi_{xx}(\infty,\kappa)$ vanishes, because the average value of the guiding
center velocity component $v_y(t)$  is finite,
while that of $v_x(t)$ is zero. 
Since $ \Phi_{xx}(\Omega \tau,\kappa)$ behaves similar to $ \Phi_{xx}(\Omega
\tau)$ in the square-symmetric case, we extrapolated Eq.~(\ref{fixxsmall}) to
arbitrary values of $\Omega \tau$ and found that Eq.~(\ref{fixx-appr})
provides an extremely good approximation.

Since, in the clean limit,  $X=1/(\Omega \tau) \rightarrow 0$, we can easily
evaluate Eq.~(\ref{w-hilf}) along an open equipotential,
 $W^-_x(-\varphi_{\epsilon},\varphi_{\epsilon};\epsilon)=-2\pi$, we
 obtain asymptotically from Eq.~(\ref{diff-open})
$\Phi_{yy}(\infty,\kappa)=\int_{\kappa-1}^{1-\kappa} d\epsilon \,4/(\Omega
T_{\epsilon} ) \,$ 
with
$$T_{\epsilon}  =2 {\bf K}(q)/(q \Omega \sqrt{\kappa}),
\quad q=(4\kappa/[(1+\kappa)^2-\epsilon^2])^{1/2} \, ,$$
where $ {\bf K}(q)$ is the complete elliptic integral \cite{Gradshteyn}.
 Expanding this for $\kappa \ll 1$,
we obtain the leading terms 
$\Phi_{yy}(\infty,\kappa) \approx 1- (31/6\pi) \kappa^{3/2}$.  Adding a
 suitable term to satisfy $\Phi_{yy}(\infty, 1)=0$, we obtained the
 approximation $\Phi_{yy}(\infty,\kappa) \approx \Phi_2(\kappa)$ as defined by
Eq.~(\ref{two-pwr-interp}).
Apparently the plot in Fig.~\ref{fig:fvonkappa} reveals only for
 $\kappa 
\gtrsim 0.7$ slight deviations between exact and interpolated result.

 Using Eq.~(\ref{two-pwr-interp}) and the small $\Omega\tau$ expansion
 (\ref{fiyysmall}), we tried to approximate $\Phi_{yy}(\Omega\tau,\kappa)$
 by the one-parameter interpolation
\begin{eqnarray}  
\Phi_{yy}^{(1)}(\Omega\tau,\kappa)=&& \Phi_{yy}(\infty,\kappa)  \nonumber \\
&& +\left[1-\Phi_{yy}(\infty,\kappa)\right] / \left[1+ \beta_\kappa
(\Omega\tau)^2 \right]\, ,  \label{fiyy_1}
\end{eqnarray}
with $\Phi_{yy}(\infty,\kappa)= \Phi_2(\kappa)$ and $ \beta_\kappa=
0.5 \kappa^2/[1-\Phi_2(\kappa)]$.
 This yields a very good approximation for
$\kappa <0.5$, but a rather poor one for $\kappa \gtrsim 0.75$, and we
improved it with the definition  (\ref{fiyy_3}).


\begin{thebibliography}{10}

\bibitem{Weiss89:179}
D. Weiss, K. von Klitzing, K. Ploog, and G. Weimann, Europhys. Lett. {\bf 8},
  179  (1989), see also in {\em High Magnetic Fields in Semiconductor Physics
  II}, edited by G. Landwehr, Springer Series in Solid-State Sciences Vol. {\bf
  87} (Springer-Verlag, Berlin 1989), p. 357.

\bibitem{Gerhardts89:1173}
R.~R. Gerhardts, D. Weiss, and K. von Klitzing, Phys. Rev. Lett. {\bf 62},
  1173  (1989).

\bibitem{Winkler89:1177}
R.~W. Winkler, J.~P. Kotthaus, and K. Ploog, Phys. Rev. Lett. {\bf 62},  1177
  (1989).

\bibitem{Weiss90:88}
D. Weiss, K. von Klitzing, K. Ploog, and G. Weimann, Surf. Sci. {\bf 229},  88
  (1990).

\bibitem{Fang90:10171}
H. Fang and P.~J. Stiles, Phys. Rev. B {\bf 41},  10171  (1990).

\bibitem{Gerhardts91:5192}
R.~R. Gerhardts, D. Weiss, and U. Wulf, Phys. Rev. B {\bf 43},  5192  (1991).

\bibitem{Lorke91:3447}
A. Lorke, J. Kotthaus, and K. Ploog, Phys. Rev. B {\bf 44},  3447  (1991).

\bibitem{Weiss92:314}
D. Weiss, A. Menschig, K. von Klitzing, , and G. Weimann, Surf. Sci. {\bf 263},
   314  (1992).

\bibitem{Zhang90:12850}
C. Zhang and R.~R. Gerhardts, Phys. Rev. B {\bf 41},  12850  (1990).

\bibitem{Hofstadter76:2239}
R.~D. Hofstadter, Phys. Rev. B {\bf 14},  2239  (1976).

\bibitem{Pfannkuche92:12606}
D. Pfannkuche and R.~R. Gerhardts, Phys. Rev. B {\bf 46},  12606  (1992).

\bibitem{Beenakker89:2020}
C.~W.~J. Beenakker, Phys. Rev. Lett. {\bf 62},  2020  (1989).

\bibitem{Zwerschke99:5536}
S.~D.~M. Zwerschke, A. Manolescu, and R.~R. Gerhardts, Phys. Rev. B {\bf 60},
  5536  (1999).

\bibitem{Gerhardts92:3449}
R.~R. Gerhardts, Phys. Rev. B {\bf 45},  3449  (1992).

\bibitem{Menne98:1707}
R. Menne and R.~R. Gerhardts, Phys. Rev. B {\bf 57},  1707  (1998).

\bibitem{Gerhardts96:11064}
R.~R. Gerhardts, Phys. Rev. B {\bf 53},  11064  (1996).

\bibitem{Schmidt93:13007}
G.~J.~O. Schmidt, Phys. Rev. B {\bf 47},  13007  (1993).

\bibitem{Grant00:13127}
D.~E. Grant, A.~R. Long, and J.~H. Davies, Phys. Rev. B {\bf 61},  13127
  (2000).

\bibitem{Chowdhury00:R4821}
S. Chowdhury, C.~J. Emeleus, B. Milton, E. Skuras, A.~R. Long, J.~H. Davies, G.
  Pennelli, and C.~R. Stanley, Phys. Rev. B {\bf 62},  R4821  (2000).

\bibitem{Fleischmann92:1367}
R. Fleischmann, T. Geisel, and R. Ketzmerick, Phys. Rev. Lett. {\bf 68},  1367
  (1992).

\bibitem{Schuster93:6843}
R. Schuster, K. Ensslin, J.~P. Kotthaus, M. Holland, and C. Stanley, Phys. Rev.
  B {\bf 47},  6843  (1993).

\bibitem{Chambers69:175}
R. G. Chambers, in {\em The Physics of Metals, I: Electrons}, edited by J. M.
  Ziman, (Cambridge University Press 1969, London), p. 175.

\bibitem{Ye96:1613}
P.~D. Ye, D. Weiss, R.~R. Gerhardts, G. L{\"u}tjering, K. von Klitzing, and H.
  Nickel, Semicond. Sci. Technol. {\bf 11},  1613  (1996).

\bibitem{Gradshteyn}
I.~S. Gradshteyn and I.~M. Ryzhik, {\em Table of Integrals, Series, and
  Products} (Academic Press, New York, 1994).

\end{thebibliography}
\end{document}